\documentclass[aps,prb,twocolumn,superscriptaddress,showpacs]{revtex4}

\usepackage{graphicx}
\usepackage{bm}
\usepackage{color}

\begin{document}


\title{Landauer theory of ballistic torkances 
             in non-collinear spin valves} 



\author{K. Carva}
\email{carva@karlov.mff.cuni.cz}
\affiliation{Charles University, Faculty of Mathematics and Physics,
  Department of Condensed Matter Physics,
  Ke Karlovu 5, CZ-12116 Prague 2, Czech Republic}
\affiliation{Uppsala University, Department of Physics 
  and Materials Science, P.O. Box 530, SE-75121 Uppsala, Sweden}
\author{I. Turek}
\email{turek@ipm.cz}
\altaffiliation[Also at ]{Institute of Physics of Materials,
  Academy of Sciences of the Czech Republic,
  CZ-61662 Brno, Czech Republic}
\affiliation{Charles University, Faculty of Mathematics and Physics,
  Department of Condensed Matter Physics,
  Ke Karlovu 5, CZ-12116 Prague 2, Czech Republic}


\date{\today}

\begin{abstract}

We present a theory of voltage-induced spin-transfer torques in 
ballistic non-collinear spin valves.
The torkance on one ferromagnetic layer is expressed in terms of
scattering coefficients of the whole spin valve, in analogy to the 
Landauer conductance formula.
The theory is applied to Co/Cu/Ni(001)-based systems where 
long-range oscillations of the Ni-torkance as a function of 
Ni thickness are predicted. 
The oscillations represent a novel quantum size effect due to
the non-collinear magnetic structure. 
The oscillatory behavior of the torkance contrasts 
a thickness-independent trend of the conductance.

\end{abstract}


\maketitle


\section{Introduction \label{s_intr}}

The prediction \cite{r_1996_jcs, r_1996_lb} and 
realization \cite{r_1999_mrk, r_2000_kab} of current-induced switching 
of magnetization direction in epitaxial magnetic multilayers 
stimulated huge research activity related to 
high-density writing of information.
The simplest systems for this purpose are spin valves 
NM/FM1/NM/FM2/NM 
with two ferromagnetic (FM) layers (FM1, FM2) separated by a 
non-magnetic (NM) spacer layer and attached to semiinfinite 
NM metallic leads.  
The electric current perpendicular to the layers becomes 
spin-polarized on passing the FM1 layer with a fixed 
magnetization direction. 
In non-collinear spin valves,  
subsequent reflection and transmission of spin-polarized 
electrons at the FM2 layer results in a spin torque acting on
its magnetization the direction of which can thus be changed. 
Majority of existing experimental and theoretical studies of 
these spin-transfer torques refer to a diffusive regime of 
electron transport in metallic systems, 
see Ref.~\onlinecite{r_2006_bbk} for a review. 

Magnetic tunnel junctions with the NM spacer layer replaced by 
an insulating barrier have attracted attention only very recently;
in these systems voltage-driven spin-transfer torques \cite{r_2007_ss} 
as well as effects of finite bias \cite{r_2006_tkk, r_2008_hs} can be 
studied. 
The concept of torkance, defined in the small-bias limit 
as a ratio of the spin-transfer torque and the applied 
voltage, \cite{r_2007_ss} represents an analogy to the conductance. 
It becomes important also for all-metallic spin valves with
ultrathin layers \cite{r_2005_efm, r_2008_wxx} where a 
ballistic regime of electron transport can be realized. 

The latter regime is amenable to fully microscopic, quantum-mechanical
treatments. 
All existing theoretical approaches to the torkance, 
both on model \cite{r_2000_wmb, r_2005_efm, r_2006_tkk, r_2007_hwd}
and \emph{ab initio} \cite{r_2008_wxx, r_2008_hs} levels, are based on 
a linear response of various local quantities inside the spin valve to 
the applied bias.
The local quantities used range from scattering coefficients of the
individual layers \cite{r_2000_wmb} over local spin 
currents \cite{r_2005_efm} to site- and orbital-resolved 
elements of a one-particle density matrix. \cite{r_2007_hwd}
These methods contrast the well-known Landauer picture of the 
ballistic conductance \cite{r_1970_rl, r_1995_sd} which employs only 
transmission coefficients between propagating states of the two 
leads. 

In this paper, we present an alternative theoretical approach
to ballistic torkances that yields a result similar to the 
Landauer conductance formula, i.e.,   
we relate the torkance to scattering coefficients of the whole 
spin valve. 
This unified theory of both transport quantities is used 
to discuss a special consequence of ballistic transport, 
namely, a predicted oscillatory dependence on Ni thickness in 
a Cu/Co/Cu/Ni/Cu(001) system. 
The presented study reveals a relation between the 
torkance and the properties of individual parts of the 
spin valve which might be relevant for design of new systems.

\section{Theory \label{s_theo}}

\subsection{Model of the spin valve \label{ss_model}}
 
Our approach is based on an effective one-electron
Hamiltonian of the NM/FM1/NM/FM2/NM system,
\begin{equation}
H = H_0 + \gamma_1 {\bf n}_1 \cdot \bm{\sigma} 
        + \gamma_2 {\bf n}_2 \cdot \bm{\sigma} , 
\label{eq_hdef}
\end{equation}
where $H_0$ comprises all spin-independent terms, 
$\gamma_1 = \gamma_1({\bf r})$  
and $\gamma_2 = \gamma_2({\bf r})$ denote magnitudes of 
exchange splittings of the FM1 and FM2 layers, respectively,
${\bf n}_1$ and ${\bf n}_2$ are unit vectors parallel to
directions of the exchange splittings, and 
the $\bm{\sigma} = ( \sigma_x, \sigma_y, \sigma_z )$ 
are the Pauli spin matrices. 
The angle between ${\bf n}_1$ and ${\bf n}_2$ is denoted 
as $\theta$. 
The spin torque $\bm{\tau}$ is defined as time derivative
of the electron spin, represented (in units of Bohr 
magneton) by operator $\bm{\sigma}$. 
This yields (with $\hbar=1$) the total spin torque as 
\begin{equation}
\bm{\tau} =  - {\rm i} \left[ \bm{\sigma} , H \right] = 
 \bm{\tau}_1 + \bm{\tau}_2 , 
\label{eq_tdef}
\end{equation}
where the quantities  
\begin{equation}
\bm{\tau}_j = 2 \gamma_j {\bf n}_j \times \bm{\sigma} ,
\qquad j = 1, 2, 
\label{eq_tjdef}
\end{equation}
can be interpreted as torques experienced by the two FM 
layers. 
Obviously, the torque $\bm{\tau}_j$ is perpendicular to
the vector ${\bf n}_j$ and it can thus be decomposed 
with respect to the common plane of the two magnetization
vectors into its in-plane ($\bm{\tau}_{j \|}$) and 
out-of-plane ($\bm{\tau}_{j \perp}$) components, see Fig.~1 
for $j=2$. 
The unit normal vector of the plane is given by 
$\bm{\nu} = {\bf n}_1 \times {\bf n}_2 / \sin \theta$.

\begin{figure}[htb]
\includegraphics[width=0.5\columnwidth]{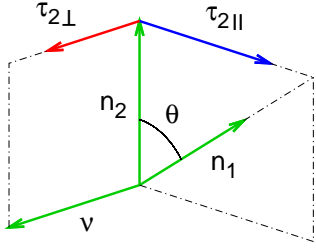}
\caption{(Color online)
The in-plane ($\bm{\tau}_{2 \|}$) and
out-of-plane ($\bm{\tau}_{2 \perp}$) components of the torque 
$\bm{\tau}_2 = \bm{\tau}_{2 \|} + \bm{\tau}_{2 \perp}$ 
experienced by the FM2 layer. 
For details, see text.}
\end{figure}

\subsection{In-plane torkance \label{ss_inplt}}

The basic idea for the in-plane torkance on the FM2 layer 
rests on the orthogonality relations 
${\bf n}_j \cdot \bm{\tau}_j = 0$, $j = 1, 2$,
from which the size of $\bm{\tau}_{2 \|}$ can be written as 
\begin{equation}
\left( {\bf n}_2 \times \bm{\nu} \right) \cdot \bm{\tau}_{2} 
= \frac{ {\bf n}_1 \cdot \bm{\tau}_2 }{ \sin \theta }
= \frac{ {\bf n}_1 \cdot \bm{\tau} }{ \sin \theta } ,
\label{eq_tau2in}
\end{equation}
see Fig.~1.
The total torque $\bm{\tau}$, being a full time derivative of 
$\bm{\sigma}$, in (\ref{eq_tau2in}) plays a key role in the
following treatment. 

Our approach applies to systems consisting of the left 
(${\cal L}$) and the right (${\cal R}$) semiinfinite NM leads 
with an intermediate region (${\cal I}$) in between; 
the latter contains both FM layers and the NM spacer of 
the spin valve. 
Projection operators on these regions are denoted respectively 
as $\Pi_{\cal L}$, $\Pi_{\cal R}$ and $\Pi_{\cal I}$; 
they are mutually orthogonal and satisfy 
$\Pi_{\cal L} + \Pi_{\cal I} + \Pi_{\cal R} = 1$.
The Hamiltonian (\ref{eq_hdef}) is assumed to be short-ranged
(tight-binding), not coupling the two leads, i.e., 
$\Pi_{\cal L} H \Pi_{\cal R} = 0$. 
The leads are in thermodynamic equilibrium at zero temperature.
A general linear response theory can be formulated for 
a Hermitean operator $Q = Q^+$ that is local, i.e., 
not coupling neighboring parts of the system, so that 
$Q = \Pi_{\cal L} Q \Pi_{\cal L} +
\Pi_{\cal I} Q \Pi_{\cal I} + \Pi_{\cal R} Q \Pi_{\cal R}$. 
Its time derivative 
\begin{equation}
D = - {\rm i} \left[ Q , H \right] 
\label{eq_ddef}
\end{equation}
is assumed to be localized in ${\cal I}$, i.e.,
$D = \Pi_{\cal I} D \Pi_{\cal I}$. 
These properties make it possible to remove 
the semiinfinite leads from the formalism. 

The resulting response coefficient describing the change
$\delta {\bar D}$ of the thermodynamic average of the 
quantity $D$ due to an infinitesimal variation 
$\delta \mu_{\cal L}$ of the chemical potential (Fermi energy)
of the ${\cal L}$ lead is given by
\begin{equation}
\frac{\delta \bar{D}}{\delta \mu_{\cal L}} 
= \frac{1}{2\pi} {\rm Tr} \left\{ 
Q \left( \Gamma_{\cal R} G^r \Gamma_{\cal L} G^a - 
 \Gamma_{\cal L} G^r \Gamma_{\cal R} G^a \right) \right\} ,
\label{eq_rcd}
\end{equation}
where the trace (Tr) and all symbols on the r.h.s.\ 
are defined on the Hilbert space of the intermediate region 
${\cal I}$, in particular the $Q$ in (\ref{eq_rcd}) abbreviates 
$\Pi_{\cal I} Q \Pi_{\cal I}$.
The other symbols in (\ref{eq_rcd}) refer to the 
antihermitean part of the ${\cal L}$ and ${\cal R}$
selfenergies, $\Gamma_{\cal L,R}(E) =
{\rm i} \left[ \Sigma^r_{\cal L,R}(E)
 - \Sigma^a_{\cal L,R}(E) \right]$, 
and to the retarded and advanced propagators 
\begin{equation}
G^{r,a}(E) = \left[ E - H - \Sigma^{r,a}(E) \right]^{-1} , 
\label{eq_gra}
\end{equation}
where $\Sigma^{r,a}(E) = \Sigma^{r,a}_{\cal L}(E) +
\Sigma^{r,a}_{\cal R}(E)$ denotes the total selfenergy.
Omitted energy arguments in (\ref{eq_rcd}) are equal to the
Fermi energy of the equilibrium system ($E = E_F$). 

The proof of (\ref{eq_rcd}) is based on non-equilibrium 
Green's functions (NGF) for stationary states \cite{r_1996_hj} 
and it is similar to a previous derivation in 
Ref.~\onlinecite{r_2007_ct}. 
The starting point is an expression for the variation of 
${\bar D}$, 
\begin{equation}
\delta {\bar D} = \frac{1}{2\pi} \int_{-\infty}^{\infty} 
{\rm Tr} \left\{ G^a(E) D G^r(E) 
\delta \Sigma^<(E) \right\} {\rm d}E , 
\label{eq_vard}
\end{equation}
where the variation of the lesser part of the selfenergy 
at zero temperature is given by 
\begin{equation}
\delta \Sigma^<(E) = \delta (E - E_F) 
\Gamma_{\cal L}(E) \delta \mu_{\cal L} .
\label{eq_vars}
\end{equation}
The assumed properties of $H$, $Q$ and $D$ lead to 
a commutation rule for the selfenergy, 
\begin{equation}
 [ Q , \Sigma^{r,a}_{\cal L,R} (E) ] = 0 , 
\label{eq_crqs}
\end{equation}
which is proved in the Appendix and
which in turn yields a relation 
\begin{eqnarray}
 G^a(E) D G^r(E) & = & 
  {\rm i} \left[ G^a(E) Q - Q G^r(E) \right] 
\nonumber\\
  & & {} + G^a(E) Q \Gamma(E) G^r(E) , 
\label{eq_gdg}
\end{eqnarray}
where
$\Gamma(E) = {\rm i} \left[ \Sigma^r(E) - \Sigma^a(E) \right] 
           = \Gamma_{\cal L}(E) + \Gamma_{\cal R}(E)$. 
The result (\ref{eq_rcd}) follows then from an 
identity for the spectral density operator, 
\begin{equation}
{\rm i} \left[ G^r(E) - G^a(E) \right] =
G^a(E) \Gamma(E) G^r(E) .
\label{eq_sdo}
\end{equation}
Note that the final response coefficient (\ref{eq_rcd}) obeys 
a perfect ${\cal L}$--${\cal R}$ symmetry, i.e., 
$\delta {\bar D} / \delta \mu_{\cal R} = 
-\delta {\bar D} / \delta \mu_{\cal L}$.

It should be emphasized that the derived general 
result (\ref{eq_rcd}) and its perfect ${\cal L}$--${\cal R}$ 
symmetry are valid only for operators $D$ that can be formulated
as a time derivative of a local operator $Q$ according to 
(\ref{eq_ddef}). 
In the present context of spin valves, this is the case of the 
usual particle conductance and of the in-plane torkance (see below). 
The out-of-plane torkance requires a different approach based on 
the more general relation (\ref{eq_vard}), 
see Section~\ref{ss_ooplt}; 
its symmetry properties for symmetric spin valves were discussed
in details, e.g., in Ref.~\onlinecite{r_2008_rs}.

Application of the derived formula (\ref{eq_rcd}) to 
the transport properties of the spin valves is now 
straightforward. 
The usual particle conductance $C$ is based on the operator 
$Q$ being a projector on a half-space containing, e.g.,
the ${\cal R}$ lead and an adjacent part of the ${\cal I}$ 
region. 
This results in the well-known expression \cite{r_1995_sd}
\begin{equation}
C = \frac{1}{2\pi} {\rm Tr} 
\left( \Gamma_{\cal R} G^r \Gamma_{\cal L} G^a \right) ,
\label{eq_cvgf}
\end{equation}
where atomic units ($e = \hbar = 1$) are used. 
The in-plane torkance $C_{\|}$ on FM2 according to 
(\ref{eq_tau2in}) is obtained from 
$Q = {\bf n}_1 \cdot \bm{\sigma}$ in (\ref{eq_rcd}). 
This yields $C_{\|} = C_1 / \sin \theta$, where 
\begin{equation}
C_1 = \frac{1}{2\pi} {\rm Tr} \left\{ 
 {\bf n}_1 \cdot \bm{\sigma} 
\left( \Gamma_{\cal R} G^r \Gamma_{\cal L} G^a - 
\Gamma_{\cal L} G^r \Gamma_{\cal R} G^a \right) \right\} .
\label{eq_tinvgf}
\end{equation}
The two terms on r.h.s.\ can be related to spin fluxes on two
sides of the FM2 layer. 
The expression (\ref{eq_tinvgf}) represents our central result.
The operators $\Gamma_{\cal L}$ and $\Gamma_{\cal R}$ are 
localized in narrow regions at the interfaces 
${\cal L}/{\cal I}$ and ${\cal I}/{\cal R}$, respectively. 
The Green's functions (propagators) for points deep inside
the spin valve thus enter neither the conductance
(\ref{eq_cvgf}), nor the in-plane torkance (\ref{eq_tinvgf}). 

\subsection{Out-of-plane torkance \label{ss_ooplt}}

A similar approach for the out-of-plane torkance on the FM2 layer
employs an infinitesimal variation $\delta {\bf n}_2$ of its 
magnetization direction due to a variation $\delta \theta$ of the 
angle. 
The FM1 magnetization direction as well as the plane of the two 
directions ${\bf n}_1, {\bf n}_2$ remain fixed, i.e., 
$\delta {\bf n}_1 = \delta \bm{\nu} = 0$. 
This leads to 
$\delta {\bf n}_2 = \bm{\nu} \times {\bf n}_2 \delta \theta$
and from (\ref{eq_hdef}) also to 
\begin{equation}
H' \equiv \frac{\delta H}{\delta \theta} = 
\gamma_2 ( \bm{\nu} \times {\bf n}_2 ) \cdot \bm{\sigma}
= \frac{1}{2}  \bm{\nu} \cdot \bm{\tau}_2 ,
\label{eq_hder}
\end{equation}
so that the size of $\bm{\tau}_{2\perp}$ coincides
(up to factor of 2) with angular derivative of the
effective Hamiltonian $H$. 
The NGF formulation of the out-of-plane torkance
rests on relation (\ref{eq_vard}) 
applied to the operator $D = 2 H'$, see (\ref{eq_hder}), 
with variation of the selfenergy $\delta \Sigma^<(E)$ due to
an infinitesimal variation of the chemical potential 
$\delta \mu_{\cal L}$ given by (\ref{eq_vars}) 
and similarly for $\delta \Sigma^<(E)$ due to the 
$\delta \mu_{\cal R}$.
This yields then response coefficients 
$C_{\cal L} = \delta \bar{D} / \delta \mu_{\cal L}$ and
$C_{\cal R} = \delta \bar{D} / \delta \mu_{\cal R}$ for 
the out-of-plane torque with respect to chemical 
potentials of the ${\cal L}$ and ${\cal R}$ leads expressed as 
\begin{equation}
C_{\cal L, R} = \frac{1}{\pi} {\rm Tr}
\left( H' G^r \Gamma_{\cal L, R} G^a \right) . 
\label{eq_clr1}
\end{equation}
By employing a simple consequence of (\ref{eq_sdo}), 
$G^a = ( 1 + {\rm i} G^a \Gamma ) G^r$, cyclic invariance
of trace, angular independence of the selfenergy 
of NM leads, $\Sigma'^r = \Sigma'^a = 0$, and the rule
$G^r H' G^r = G'^r$, the response coefficients can be recast into 
\begin{equation}
C_{\cal L, R} = \frac{1}{\pi} {\rm Tr} \left\{
G'^r \Gamma_{\cal L, R} \left[ 1 + {\rm i} G^a
( \Gamma_{\cal L} + \Gamma_{\cal R} ) \right] \right\} , 
\label{eq_clr2}
\end{equation}
which contain again only propagators at points close to the
${\cal L}/{\cal I}$ and ${\cal I}/{\cal R}$ interfaces, 
similarly to (\ref{eq_cvgf}, \ref{eq_tinvgf}).

The applied bias has to be identified with the difference
$\mu_{\cal L} - \mu_{\cal R}$ and the out-of-plane torkance on FM2 
is thus given by $C_{\perp} = ( C_{\cal L} - C_{\cal R} )/2$. 
Since the Hamiltonian (\ref{eq_hdef}) does not contain spin-orbit
interaction, the spin reference system can be chosen such that 
both unit vectors ${\bf n}_1$ and ${\bf n}_2$ lie in the $x-z$ 
plane.
This implies that $H$ is essentially time-inversion invariant and 
it can be represented by a symmetric matrix, $H^T = H$; 
the related quantities $G^{r,a}$ and $\Gamma_{\cal L, R}$ 
are symmetric as well. 
As a consequence, the transmission-like terms in 
$C_{\cal L}$ and $C_{\cal R}$ are the same, i.e., 
${\rm Tr} \left( G'^r \Gamma_{\cal L} G^a \Gamma_{\cal R} \right) =
 {\rm Tr} \left( G'^r \Gamma_{\cal R} G^a \Gamma_{\cal L} \right)$. 
The resulting out-of-plane torkance 
\begin{equation}
C_{\perp} = \frac{1}{2\pi} {\rm Tr} \left\{
G'^r [ \Gamma_{\cal L} ( 1 + {\rm i} G^a \Gamma_{\cal L} ) 
- \Gamma_{\cal R} ( 1 + {\rm i} G^a \Gamma_{\cal R} ) ] \right\}
\label{eq_touvgf}
\end{equation}
contains thus only reflection-like terms. 

\subsection{Landauer formalism \label{ss_landf}}

The Green's function expression for the conductance (\ref{eq_cvgf}) 
can be translated in the language of scattering 
theory; \cite{r_2006_am}
the counterparts of the torkances (\ref{eq_tinvgf}, \ref{eq_touvgf})
are interesting as well.  
In the present case, propagating states in the ${\cal L}$ and 
${\cal R}$ lead will be labelled by $\lambda$ and $\rho$, 
respectively. 
Moreover, a spin index $s = \uparrow, \downarrow$ has to be used 
even for NM leads, since non-collinearity of the spin valve 
gives rise to full spin dependence of scattering coefficients.  

The conductance (\ref{eq_cvgf}) is given by the Landauer formula
$C = (2\pi)^{-1} \sum_{\lambda \rho s s'} 
\left| t_{\rho s', \lambda s} \right|^2$, 
where $t_{\rho s', \lambda s}$ denotes the transmission coefficient
from an incoming state $\lambda s$ into 
an outgoing state $\rho s'$. \cite{r_1970_rl} 
The in-plane torkance coefficient (\ref{eq_tinvgf}) can be 
written as 
\begin{equation}
C_1 = \frac{1}{2\pi} \sum_{\lambda \rho s s' s''}
( {\bf n}_1 \cdot \bm{\sigma} )_{s''s'}
 \left( t_{\rho s', \lambda s} t^{\ast}_{\rho s'', \lambda s}
 - t_{\lambda s', \rho s} t^{\ast}_{\lambda s'', \rho s} \right) ,
\label{eq_lftin} 
\end{equation}
whereas the out-of-plane torkance (\ref{eq_touvgf}) can be 
transformed into
\begin{equation}
C_{\perp} = \frac{\rm i}{2\pi} \left( \sum_{\lambda s \lambda' s'}
r'_{\lambda' s', \lambda s} r^{\ast}_{\lambda' s', \lambda s} 
 - \sum_{\rho s \rho' s'} 
r'_{\rho' s', \rho s} r^{\ast}_{\rho' s', \rho s} \right) ,  
\label{eq_lftout} 
\end{equation}
where $r_{\lambda' s', \lambda s}$ ($r_{\rho' s', \rho s}$) 
denote reflection coefficients between states of the ${\cal L}$ 
(${\cal R}$) lead.
This result represents analogy to the Landauer formula and it
completes the unified theory of conductances and torkances. 

\section{Results for Cu/Co/Cu/Ni/Cu(001) 
         and their discussion \label{s_resu}}

The developed formalism allows to study properties of spin
valves with ultrathin layers, which is yet an experimentally 
unexplored area;
here we demonstrate its use for understanding unexpected features 
of \emph{ab initio} results. 
The results discussed below were obtained using the response of 
spin currents on both sides of the FM2 layer, 
\cite{r_2005_efm, r_2006_tkk} implemented within the scalar relativistic 
tight-binding linear muffin-tin orbital (TB-LMTO) 
method \cite{r_1984_aj, r_1997_tdk} similarly to 
our previous transport studies. \cite{r_2006_ctk, r_2007_ct}
As a case study, spin valves Cu/Co/Cu/Ni/Cu(001) with 
face-centered cubic (fcc) structure were chosen. 
All atomic positions were given by an ideal fcc Co lattice with
sharp interfaces between the neighboring FM and NM layers. 
The spin valves discussed below consist of a Co layer of 
5 monolayer (ML) thickness separated by a 10 ML thick Cu spacer 
from a Ni layer of varying thickness, 
embedded between two semiinfinite Cu leads. 
Self-consistent calculations within the local spin-density
approximation (LSDA) were performed only for collinear spin
valves ($\theta = 0$ or $\theta = \pi$) while the electronic
structure of non-collinear systems was obtained by rotation
of the exchange-split potentials of the Co and Ni FM layers. 
Particular attention has been paid to the convergence of 
torkances with respect to the number of ${\bf k}_\|$ vectors
sampling the two-dimensional Brillouin zone (BZ) of the system;
in agreement with Ref.~\onlinecite{r_2008_hcy} we found that
reliable values of the out-of-plane torkances require finer 
meshes than for the in-plane torkances.
The presented data were obtained with 6400 ${\bf k}_\|$ points 
in the whole BZ.

\begin{figure}[htb]
\includegraphics[width=0.8\columnwidth]{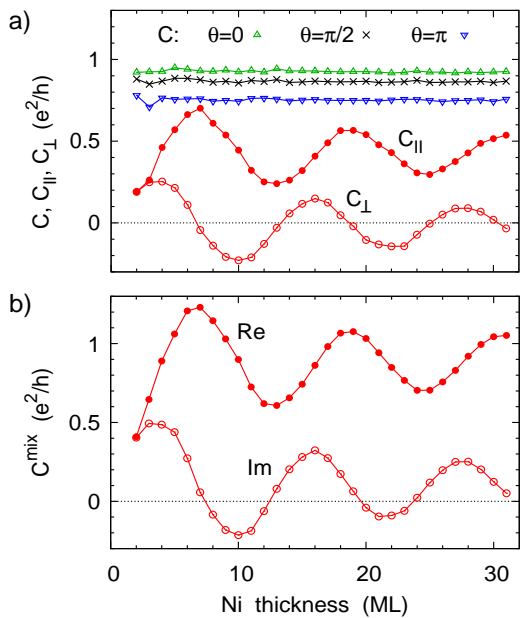}
\caption{(Color online)
Calculated transport coefficients (per interface atom) 
as functions of Ni thickness:
(a) the conductance ($C$) for three values of the angle $\theta$ 
and the in-plane ($C_{\|}$) and out-of-plane ($C_{\perp}$) 
Ni-torkances for $\theta = \pi/2$ in spin valves 
Cu/Co/Cu/Ni/Cu(001), 
(b) the real (Re) and imaginary (Im) parts of the spin-mixing 
conductance ($C^{\rm mix}$) of Cu/Ni/Cu(001) systems.}
\end{figure}

Figure 2a displays the calculated conductances for parallel 
($\theta = 0$), antiparallel ($\theta = \pi$) and perpendicular 
($\theta = \pi/2$) orientations as well as Ni-torkances in the 
latter case as functions of Ni thickness.  
The most pronounced feature of the transport coefficients are 
oscillations with a period of about 12 ML seen in both components
of the torkance. 
These oscillations reflect the perfect ballistic regime of 
electron transport across the whole spin valve. 
In addition, they contradict a generally accepted idea of very 
short magnetic coherence lengths of a few interatomic spacings, or, 
equivalently, of the spin-transfer torques as an interface 
property. \cite{r_2006_bbk, r_2007_hwd, r_2008_hs, r_2002_sz}
Very recently, spin-transfer torques in antiferromagnetic metallic
FeMn layers have been investigated theoretically; \cite{r_2008_xwx}
it has been shown that the torques are not localized to the
interface but are effective over the whole FeMn layer. 
However, no oscillatory behavior of the total torkance as 
a function of the layer thickness has been reported.
The nature of the predicted oscillations deserves thus detailed 
analysis, including also a discussion of their 
stability with respect to structural imperfections and of their 
absence in the conductance (see Fig.~2a). 

Oscillations similar to those in Fig.~2a have recently been obtained 
for a different quantity of a simpler system, namely for the 
spin-mixing conductance $C^{\rm mix}$ of epitaxial fcc (001) Ni thin 
films attached to Cu leads. \cite{r_2007_ct} 
The real and imaginary parts of the complex $C^{\rm mix}$ are related
to two components of the spin torque experienced by the FM film due 
to a spin accumulation in one of the NM leads; \cite{r_2006_bbk} the 
calculated values of $C^{\rm mix}$ for the Cu/Ni/Cu(001) system are 
shown in Fig.~2b. 
The oscillation periods of the torkance and the spin-mixing 
conductance are identical which indicates a common origin of both. 
The physical mechanism behind the $C^{\rm mix}$ oscillations was 
identified with an interference effect between spin-$\uparrow$ 
electrons propagating across the Ni film from the ${\cal R}$ lead 
to the ${\cal L}$ lead and spin-$\downarrow$ electrons propagating 
backwards. 
This effect is expressed by a spin-mixing term 
$\sim {\rm tr} ( \Gamma_{\cal R} G^r_\uparrow 
\Gamma_{\cal L} G^a_\downarrow ) $ in the $C^{\rm mix}$, where
the $G^{r,a}_s$ ($s = \uparrow, \downarrow$) denote spin-resolved 
propagators and the trace (tr) does not involve the spin 
index. \cite{r_2007_ct} 
The particular value of the oscillation period follows from a 
special shape of the spin-polarized Fermi surface of bulk fcc 
Ni. \cite{r_2007_ct}

The oscillations of $C^{\rm mix}$ have been found fairly stable 
against Cu-Ni interdiffusion at the interfaces; \cite{r_2008_ct} 
the same stability can be thus expected for the torkance 
oscillations in the spin valve. 
The relative stability can be understood as an effect of the 
large oscillation period ($\sim$ 12 ML): 
intermixing confined to a very few atomic planes at interfaces 
reduces the oscillation amplitude rather weakly. 
This feature contrasts, e.g., sensitivity of the interlayer 
exchange coupling in magnetic multilayers mediated by 
a NM Cu(001) spacer with oscillation periods of 
$\sim$ 2.5 ML and 6 ML, where even a very small amount of interface 
disorder reduces strongly especially the amplitude of the 
short period oscillations. \cite{r_1996_kdt} 

\begin{figure}[htb]
\includegraphics[width=0.8\columnwidth]{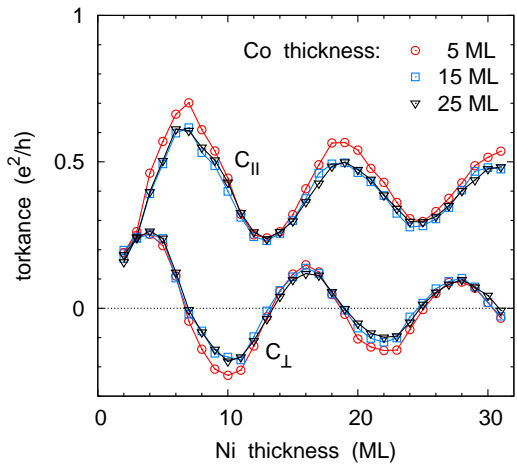}
\caption{(Color online)
Calculated in-plane ($C_{\|}$) and out-of-plane ($C_{\perp}$) 
Ni-torkances (per interface atom) as functions of Ni thickness 
in spin valves Cu/Co/Cu/Ni/Cu(001) for $\theta = \pi/2$, 
10 ML Cu spacer and for three different Co thicknesses.} 
\end{figure}

Another important aspect of the oscillations of the
spin-transfer torques concerns their dependence on the thickness 
of the polarizing Co layer since ultrathin layers in general might 
amplify ballistic and interference effects. 
Figure 3 presents the Ni-torkances in the same Cu/Co/Cu/Ni/Cu(001) 
spin valves calculated for three different Co thicknesses, 
namely 5, 15 and 25 ML. 
It can be seen that the oscillations persist and have the same 
period in all three cases. 
Their amplitudes depend slightly on the Co thickness: 
the initial increase from 5 to 15 Co ML is accompanied by a small 
reduction (of about 20 \%) of the amplitudes whereas further increase 
from 15 to 25 Co ML does not influence them appreciably.
More detailed investigation of the effect of the thickness of the 
polarizing Co layer, including also the limiting case of spin valves 
FM1/NM/FM2/NM with a semiinfinite polarizing FM 
lead, \cite{r_2005_efm} is beyond the scope of the present study. 

Let us now discuss the absence of oscillations in the 
conductance (Fig.~2a). 
We introduce propagators $G^{r,a}_2$ of an auxiliary system 
NM/NM1/NM/FM2/NM, where NM1 denotes the FM1 layer with null
exchange splitting. 
These propagators satisfy 
$G^{r,a} = G^{r,a}_2 + G^{r,a}_2 T^{r,a}_1 G^{r,a}_2$ 
where the $T^{r,a}_1$ denotes the $t$-matrix corresponding
to the FM1 exchange splitting $\gamma_1$ in (\ref{eq_hdef}). 
The conductance (\ref{eq_cvgf}) can be then rewritten as
$C = (2\pi)^{-1} {\rm Tr} 
( \Gamma_{\cal R} G^r_2 \Delta_{\cal L} G^a_2 )$,
where $\Delta_{\cal L} = (1 + T^r_1 G^r_2)
\Gamma_{\cal L} (1 + G^a_2 T^a_1)$ represents an operator 
localized at the FM1 layer, i.e., at the left end of the 
FM2 layer. 
The latter trace can be most easily evaluated using the 
spin quantization axis parallel to the FM2 magnetization 
direction ${\bf n}_2$. 
Since the propagators $G^{r,a}_2$ are now diagonal in the spin 
index $s$ and the operator $\Gamma_{\cal R}$ is spin-independent, 
the conductance does not contain spin-mixing terms, i.e., 
terms $\sim {\rm tr} ( \Gamma_{\cal R} G^r_{2,s}
\Delta_{{\cal L},ss'} G^a_{2,s'} )$ for $s \ne s'$ that 
result in interference effects involving different spin channels. 
For the torkance, however, 
the extra factor ${\bf n}_1 \cdot \bm{\sigma}$ 
in (\ref{eq_tinvgf}) provides the necessary spin mixing 
responsible for the oscillations, in full analogy 
to oscillations of the spin-mixing conductance. 

A recent study of spin-transfer torques in a tunnel junction
Cu/Fe/MgO/Fe/Cu has predicted torkance and conductance 
oscillations with Fe thickness with a period 
$\sim$ 2 ML. \cite{r_2008_hs} 
These oscillations were ascribed to quantum well states 
in the majority spin of the Fe layer, i.e., to interference 
effects in a single spin channel.
The oscillations in the Cu/Co/Cu/Ni/Cu system---manifested
only in the torkance---have thus a clearly different origin.

\section{Conclusion \label{s_concl}}

We have addressed two important aspects of spin-transfer 
torques in non-collinear spin valves with ultrathin layers.
First, we have shown that the in-plane and out-of-plane
torkance on one FM layer can be expressed by means of
the transmission and reflection coefficients, respectively,
of the whole spin valve, in close analogy to the Landauer 
formula for the ballistic conductance. 
Second, a novel oscillatory behavior for Ni-based systems 
has been predicted due to the mixed spin channels. 
The oscillations with Ni thickness are reasonably stable 
with respect to interface imperfections of real samples; 
however, they are not present in the conductance but can 
be observed only in the Ni-torkance. 
The torkance oscillations prove that the spin-transfer 
torques in ballistic spin valves are closely related to 
properties of their components, in particular to the 
spin-mixing conductances of individual ferromagnetic layers.  

\begin{acknowledgments}

This work has been supported by the Ministry of Education of the 
Czech Republic (No.\ MSM0021620834) and by
the Academy of Sciences of the Czech Republic (No.\ KJB101120803, 
No.\ KAN400100653). 

\end{acknowledgments}

\appendix*
\section{Proof of the commutation rule for selfenergy} 

The proof of the commutation rule (\ref{eq_crqs}) rests on 
assumed properties of the operators $H$, $Q$, $D$ (defined in 
the Hilbert space of the total system) and of 
the projectors $\Pi_{\cal L}$, $\Pi_{\cal R}$, $\Pi_{\cal I}$, 
see the beginning of Section \ref{ss_inplt}. 
Let as abbreviate projections of any operator $X$ ($X=H,Q,D$) as 
$\Pi_{\cal I} X \Pi_{\cal I} \equiv X_{\cal I I}$, 
$\Pi_{\cal I} X \Pi_{\cal L} \equiv X_{\cal I L}$, etc.
The assumed property of $Q$, namely $Q = Q_{\cal L L} + 
Q_{\cal I I} + Q_{\cal R R}$, a consequence of the
localization of $D$ in ${\cal I}$, namely $D_{\cal I L} = 0$ and 
$D_{\cal L I} = 0$, and the orthogonality of the projectors 
$\Pi_{\cal L}$, $\Pi_{\cal R}$, $\Pi_{\cal I}$ lead to identities
\begin{equation}
Q_{\cal I I} H_{\cal I L} = 
H_{\cal I L} Q_{\cal L L} , \quad  
Q_{\cal L L} H_{\cal L I} = 
H_{\cal L I} Q_{\cal I I} . 
\label{eq_qhil}
\end{equation}
Similarly, a commutation rule 
\begin{equation}
Q_{\cal L L} H_{\cal L L} = 
H_{\cal L L} Q_{\cal L L}   
\label{eq_qhll}
\end{equation}
can easily be obtained from $D_{\cal L L} = 0$. 

The left selfenergy is given explicitly by
\begin{equation}
 \Sigma^{r,a}_{\cal L} (E) =  
H_{\cal I L} {\cal G}^{r,a}_{\cal L}(E) H_{\cal L I} , 
\label{eq_lse}
\end{equation}
where the ${\cal G}^{r,a}_{\cal L}(E)$ denotes 
the retarded and advanced propagator of the isolated left lead. 
The relation (\ref{eq_qhll}) implies immediately 
a commutation rule 
\begin{equation}
Q_{\cal L L} {\cal G}^{r,a}_{\cal L}(E) = 
{\cal G}^{r,a}_{\cal L}(E) Q_{\cal L L} ; 
\label{eq_qgl}
\end{equation}
its application together with (\ref{eq_qhil}, \ref{eq_lse})
leads to identities
\begin{eqnarray}
\lefteqn{Q_{\cal I I} 
 \Sigma^{r,a}_{\cal L} (E) = 
Q_{\cal I I} H_{\cal I L} 
{\cal G}^{r,a}_{\cal L}(E) H_{\cal L I}}  
\nonumber\\
 & = & 
H_{\cal I L} Q_{\cal L L} 
{\cal G}^{r,a}_{\cal L}(E) H_{\cal L I}  
  =  
H_{\cal I L} {\cal G}^{r,a}_{\cal L}(E) 
Q_{\cal L L} H_{\cal L I}  
\nonumber\\
 & = & 
H_{\cal I L} {\cal G}^{r,a}_{\cal L}(E) 
H_{\cal L I}  Q_{\cal I I} 
 = 
 \Sigma^{r,a}_{\cal L} (E) Q_{\cal I I} ,
\label{eq_prx}
\end{eqnarray}
which are equivalent to the commutation rule (\ref{eq_crqs}) 
for the left selfenergy.
The proof for the right selfenergy is similar and 
therefore omitted.


\begin{thebibliography}{26}
\expandafter\ifx\csname natexlab\endcsname\relax\def\natexlab#1{#1}\fi
\expandafter\ifx\csname bibnamefont\endcsname\relax
  \def\bibnamefont#1{#1}\fi
\expandafter\ifx\csname bibfnamefont\endcsname\relax
  \def\bibfnamefont#1{#1}\fi
\expandafter\ifx\csname citenamefont\endcsname\relax
  \def\citenamefont#1{#1}\fi
\expandafter\ifx\csname url\endcsname\relax
  \def\url#1{\texttt{#1}}\fi
\expandafter\ifx\csname urlprefix\endcsname\relax\def\urlprefix{URL }\fi
\providecommand{\bibinfo}[2]{#2}
\providecommand{\eprint}[2][]{\url{#2}}

\bibitem[{\citenamefont{Slonczewski}(1996)}]{r_1996_jcs}
\bibinfo{author}{\bibfnamefont{J.~C.} \bibnamefont{Slonczewski}},
  \bibinfo{journal}{J. Magn. Magn. Mater.} \textbf{\bibinfo{volume}{159}},
  \bibinfo{pages}{L1} (\bibinfo{year}{1996}).

\bibitem[{\citenamefont{Berger}(1996)}]{r_1996_lb}
\bibinfo{author}{\bibfnamefont{L.}~\bibnamefont{Berger}},
  \bibinfo{journal}{Phys. Rev. B} \textbf{\bibinfo{volume}{54}},
  \bibinfo{pages}{9353} (\bibinfo{year}{1996}).

\bibitem[{\citenamefont{Myers et~al.}(1999)\citenamefont{Myers, Ralph, Katine,
  Louie, and Buhrman}}]{r_1999_mrk}
\bibinfo{author}{\bibfnamefont{E.~B.} \bibnamefont{Myers}},
  \bibinfo{author}{\bibfnamefont{D.~C.} \bibnamefont{Ralph}},
  \bibinfo{author}{\bibfnamefont{J.~A.} \bibnamefont{Katine}},
  \bibinfo{author}{\bibfnamefont{R.~N.} \bibnamefont{Louie}}, \bibnamefont{and}
  \bibinfo{author}{\bibfnamefont{R.~A.} \bibnamefont{Buhrman}},
  \bibinfo{journal}{Science} \textbf{\bibinfo{volume}{285}},
  \bibinfo{pages}{867} (\bibinfo{year}{1999}).

\bibitem[{\citenamefont{Katine et~al.}(2000)\citenamefont{Katine, Albert,
  Buhrman, Myers, and Ralph}}]{r_2000_kab}
\bibinfo{author}{\bibfnamefont{J.~A.} \bibnamefont{Katine}},
  \bibinfo{author}{\bibfnamefont{F.~J.} \bibnamefont{Albert}},
  \bibinfo{author}{\bibfnamefont{R.~A.} \bibnamefont{Buhrman}},
  \bibinfo{author}{\bibfnamefont{E.~B.} \bibnamefont{Myers}}, \bibnamefont{and}
  \bibinfo{author}{\bibfnamefont{D.~C.} \bibnamefont{Ralph}},
  \bibinfo{journal}{Phys. Rev. Lett.} \textbf{\bibinfo{volume}{84}},
  \bibinfo{pages}{3149} (\bibinfo{year}{2000}).

\bibitem[{\citenamefont{Brataas et~al.}(2006)\citenamefont{Brataas, Bauer, and
  Kelly}}]{r_2006_bbk}
\bibinfo{author}{\bibfnamefont{A.}~\bibnamefont{Brataas}},
  \bibinfo{author}{\bibfnamefont{G.~E.~W.} \bibnamefont{Bauer}},
  \bibnamefont{and} \bibinfo{author}{\bibfnamefont{P.~J.} \bibnamefont{Kelly}},
  \bibinfo{journal}{Phys. Rep.} \textbf{\bibinfo{volume}{427}},
  \bibinfo{pages}{157} (\bibinfo{year}{2006}).

\bibitem[{\citenamefont{Slonczewski and Sun}(2007)}]{r_2007_ss}
\bibinfo{author}{\bibfnamefont{J.~C.} \bibnamefont{Slonczewski}}
  \bibnamefont{and} \bibinfo{author}{\bibfnamefont{J.~Z.} \bibnamefont{Sun}},
  \bibinfo{journal}{J. Magn. Magn. Mater.} \textbf{\bibinfo{volume}{310}},
  \bibinfo{pages}{169} (\bibinfo{year}{2007}).

\bibitem[{\citenamefont{Theodonis et~al.}(2006)\citenamefont{Theodonis,
  Kioussis, Kalitsov, Chshiev, and Butler}}]{r_2006_tkk}
\bibinfo{author}{\bibfnamefont{I.}~\bibnamefont{Theodonis}},
  \bibinfo{author}{\bibfnamefont{N.}~\bibnamefont{Kioussis}},
  \bibinfo{author}{\bibfnamefont{A.}~\bibnamefont{Kalitsov}},
  \bibinfo{author}{\bibfnamefont{M.}~\bibnamefont{Chshiev}}, \bibnamefont{and}
  \bibinfo{author}{\bibfnamefont{W.~H.} \bibnamefont{Butler}},
  \bibinfo{journal}{Phys. Rev. Lett.} \textbf{\bibinfo{volume}{97}},
  \bibinfo{pages}{237205} (\bibinfo{year}{2006}).

\bibitem[{\citenamefont{Heiliger and Stiles}(2008)}]{r_2008_hs}
\bibinfo{author}{\bibfnamefont{C.}~\bibnamefont{Heiliger}} \bibnamefont{and}
  \bibinfo{author}{\bibfnamefont{M.~D.} \bibnamefont{Stiles}},
  \bibinfo{journal}{Phys. Rev. Lett.} \textbf{\bibinfo{volume}{100}},
  \bibinfo{pages}{186805} (\bibinfo{year}{2008}).

\bibitem[{\citenamefont{Edwards et~al.}(2005)\citenamefont{Edwards, Federici,
  Mathon, and Umerski}}]{r_2005_efm}
\bibinfo{author}{\bibfnamefont{D.~M.} \bibnamefont{Edwards}},
  \bibinfo{author}{\bibfnamefont{F.}~\bibnamefont{Federici}},
  \bibinfo{author}{\bibfnamefont{J.}~\bibnamefont{Mathon}}, \bibnamefont{and}
  \bibinfo{author}{\bibfnamefont{A.}~\bibnamefont{Umerski}},
  \bibinfo{journal}{Phys. Rev. B} \textbf{\bibinfo{volume}{71}},
  \bibinfo{pages}{054407} (\bibinfo{year}{2005}).

\bibitem[{\citenamefont{Wang et~al.}(2008)\citenamefont{Wang, Xu, and
  Xia}}]{r_2008_wxx}
\bibinfo{author}{\bibfnamefont{S.}~\bibnamefont{Wang}},
  \bibinfo{author}{\bibfnamefont{Y.}~\bibnamefont{Xu}}, \bibnamefont{and}
  \bibinfo{author}{\bibfnamefont{K.}~\bibnamefont{Xia}},
  \bibinfo{journal}{Phys. Rev. B} \textbf{\bibinfo{volume}{77}},
  \bibinfo{pages}{184430} (\bibinfo{year}{2008}).

\bibitem[{\citenamefont{Haney et~al.}(2007)\citenamefont{Haney, Waldron, Duine,
  N\'u{\~n}ez, Guo, and MacDonald}}]{r_2007_hwd}
\bibinfo{author}{\bibfnamefont{P.~M.} \bibnamefont{Haney}},
  \bibinfo{author}{\bibfnamefont{D.}~\bibnamefont{Waldron}},
  \bibinfo{author}{\bibfnamefont{R.~A.} \bibnamefont{Duine}},
  \bibinfo{author}{\bibfnamefont{A.~S.} \bibnamefont{N\'u{\~n}ez}},
  \bibinfo{author}{\bibfnamefont{H.}~\bibnamefont{Guo}}, \bibnamefont{and}
  \bibinfo{author}{\bibfnamefont{A.~H.} \bibnamefont{MacDonald}},
  \bibinfo{journal}{Phys. Rev. B} \textbf{\bibinfo{volume}{76}},
  \bibinfo{pages}{024404} (\bibinfo{year}{2007}).

\bibitem[{\citenamefont{Waintal et~al.}(2000)\citenamefont{Waintal, Myers,
  Brouwer, and Ralph}}]{r_2000_wmb}
\bibinfo{author}{\bibfnamefont{X.}~\bibnamefont{Waintal}},
  \bibinfo{author}{\bibfnamefont{E.~B.} \bibnamefont{Myers}},
  \bibinfo{author}{\bibfnamefont{P.~W.} \bibnamefont{Brouwer}},
  \bibnamefont{and} \bibinfo{author}{\bibfnamefont{D.~C.} \bibnamefont{Ralph}},
  \bibinfo{journal}{Phys. Rev. B} \textbf{\bibinfo{volume}{62}},
  \bibinfo{pages}{12317} (\bibinfo{year}{2000}).

\bibitem[{\citenamefont{Landauer}(1970)}]{r_1970_rl}
\bibinfo{author}{\bibfnamefont{R.}~\bibnamefont{Landauer}},
  \bibinfo{journal}{Philosophical Magazine} \textbf{\bibinfo{volume}{21}},
  \bibinfo{pages}{863} (\bibinfo{year}{1970}).

\bibitem[{\citenamefont{Datta}(1995)}]{r_1995_sd}
\bibinfo{author}{\bibfnamefont{S.}~\bibnamefont{Datta}},
  \emph{\bibinfo{title}{Electronic Transport in Mesoscopic Systems}}
  (\bibinfo{publisher}{Cambridge University Press}, \bibinfo{year}{1995}).

\bibitem[{\citenamefont{Haug and Jauho}(1996)}]{r_1996_hj}
\bibinfo{author}{\bibfnamefont{H.}~\bibnamefont{Haug}} \bibnamefont{and}
  \bibinfo{author}{\bibfnamefont{A.-P.} \bibnamefont{Jauho}},
  \emph{\bibinfo{title}{Quantum Kinetics in Transport and Optics of
  Semiconductors}} (\bibinfo{publisher}{Springer, Berlin},
  \bibinfo{year}{1996}).

\bibitem[{\citenamefont{Carva and Turek}(2007)}]{r_2007_ct}
\bibinfo{author}{\bibfnamefont{K.}~\bibnamefont{Carva}} \bibnamefont{and}
  \bibinfo{author}{\bibfnamefont{I.}~\bibnamefont{Turek}},
  \bibinfo{journal}{Phys. Rev. B} \textbf{\bibinfo{volume}{76}},
  \bibinfo{pages}{104409} (\bibinfo{year}{2007}).

\bibitem[{\citenamefont{Ralph and Stiles}(2008)}]{r_2008_rs}
\bibinfo{author}{\bibfnamefont{D.~C.} \bibnamefont{Ralph}} \bibnamefont{and}
  \bibinfo{author}{\bibfnamefont{M.~D.} \bibnamefont{Stiles}},
  \bibinfo{journal}{J. Magn. Magn. Mater.} \textbf{\bibinfo{volume}{320}},
  \bibinfo{pages}{1190} (\bibinfo{year}{2008}).

\bibitem[{\citenamefont{Arrachea and Moskalets}(2006)}]{r_2006_am}
\bibinfo{author}{\bibfnamefont{L.}~\bibnamefont{Arrachea}} \bibnamefont{and}
  \bibinfo{author}{\bibfnamefont{M.}~\bibnamefont{Moskalets}},
  \bibinfo{journal}{Phys. Rev. B} \textbf{\bibinfo{volume}{74}},
  \bibinfo{pages}{245322} (\bibinfo{year}{2006}).

\bibitem[{\citenamefont{Andersen and Jepsen}(1984)}]{r_1984_aj}
\bibinfo{author}{\bibfnamefont{O.~K.} \bibnamefont{Andersen}} \bibnamefont{and}
  \bibinfo{author}{\bibfnamefont{O.}~\bibnamefont{Jepsen}},
  \bibinfo{journal}{Phys. Rev. Lett.} \textbf{\bibinfo{volume}{53}},
  \bibinfo{pages}{2571} (\bibinfo{year}{1984}).

\bibitem[{\citenamefont{Turek et~al.}(1997)\citenamefont{Turek, Drchal,
  Kudrnovsk\'y, \v{S}ob, and Weinberger}}]{r_1997_tdk}
\bibinfo{author}{\bibfnamefont{I.}~\bibnamefont{Turek}},
  \bibinfo{author}{\bibfnamefont{V.}~\bibnamefont{Drchal}},
  \bibinfo{author}{\bibfnamefont{J.}~\bibnamefont{Kudrnovsk\'y}},
  \bibinfo{author}{\bibfnamefont{M.}~\bibnamefont{\v{S}ob}}, \bibnamefont{and}
  \bibinfo{author}{\bibfnamefont{P.}~\bibnamefont{Weinberger}},
  \emph{\bibinfo{title}{Electronic Structure of Disordered Alloys, Surfaces and
  Interfaces}} (\bibinfo{publisher}{Kluwer, Boston}, \bibinfo{year}{1997}).

\bibitem[{\citenamefont{Carva et~al.}(2006)\citenamefont{Carva, Turek,
  Kudrnovsk\'y, and Bengone}}]{r_2006_ctk}
\bibinfo{author}{\bibfnamefont{K.}~\bibnamefont{Carva}},
  \bibinfo{author}{\bibfnamefont{I.}~\bibnamefont{Turek}},
  \bibinfo{author}{\bibfnamefont{J.}~\bibnamefont{Kudrnovsk\'y}},
  \bibnamefont{and} \bibinfo{author}{\bibfnamefont{O.}~\bibnamefont{Bengone}},
  \bibinfo{journal}{Phys. Rev. B} \textbf{\bibinfo{volume}{73}},
  \bibinfo{pages}{144421} (\bibinfo{year}{2006}).

\bibitem[{\citenamefont{Heiliger et~al.}(2008)\citenamefont{Heiliger, Czerner,
  Yavorsky, Mertig, and Stiles}}]{r_2008_hcy}
\bibinfo{author}{\bibfnamefont{C.}~\bibnamefont{Heiliger}},
  \bibinfo{author}{\bibfnamefont{M.}~\bibnamefont{Czerner}},
  \bibinfo{author}{\bibfnamefont{B.~Y.} \bibnamefont{Yavorsky}},
  \bibinfo{author}{\bibfnamefont{I.}~\bibnamefont{Mertig}}, \bibnamefont{and}
  \bibinfo{author}{\bibfnamefont{M.~D.} \bibnamefont{Stiles}},
  \bibinfo{journal}{J. Appl. Phys.} \textbf{\bibinfo{volume}{103}},
  \bibinfo{pages}{07A709} (\bibinfo{year}{2008}).

\bibitem[{\citenamefont{Stiles and Zangwill}(2002)}]{r_2002_sz}
\bibinfo{author}{\bibfnamefont{M.~D.} \bibnamefont{Stiles}} \bibnamefont{and}
  \bibinfo{author}{\bibfnamefont{A.}~\bibnamefont{Zangwill}},
  \bibinfo{journal}{Phys. Rev. B} \textbf{\bibinfo{volume}{66}},
  \bibinfo{pages}{014407} (\bibinfo{year}{2002}).

\bibitem[{\citenamefont{Xu et~al.}(2008)\citenamefont{Xu, Wang, and
  Xia}}]{r_2008_xwx}
\bibinfo{author}{\bibfnamefont{Y.}~\bibnamefont{Xu}},
  \bibinfo{author}{\bibfnamefont{S.}~\bibnamefont{Wang}}, \bibnamefont{and}
  \bibinfo{author}{\bibfnamefont{K.}~\bibnamefont{Xia}},
  \bibinfo{journal}{Phys. Rev. Lett.} \textbf{\bibinfo{volume}{100}},
  \bibinfo{pages}{226602} (\bibinfo{year}{2008}).

\bibitem[{\citenamefont{Carva and Turek}(2008)}]{r_2008_ct}
\bibinfo{author}{\bibfnamefont{K.}~\bibnamefont{Carva}} \bibnamefont{and}
  \bibinfo{author}{\bibfnamefont{I.}~\bibnamefont{Turek}},
  \bibinfo{journal}{Physica Status Solidi A} \textbf{\bibinfo{volume}{205}},
  \bibinfo{pages}{1805} (\bibinfo{year}{2008}).

\bibitem[{\citenamefont{Kudrnovsk\'y et~al.}(1996)\citenamefont{Kudrnovsk\'y,
  Drchal, Turek, \v{S}ob, and Weinberger}}]{r_1996_kdt}
\bibinfo{author}{\bibfnamefont{J.}~\bibnamefont{Kudrnovsk\'y}},
  \bibinfo{author}{\bibfnamefont{V.}~\bibnamefont{Drchal}},
  \bibinfo{author}{\bibfnamefont{I.}~\bibnamefont{Turek}},
  \bibinfo{author}{\bibfnamefont{M.}~\bibnamefont{\v{S}ob}}, \bibnamefont{and}
  \bibinfo{author}{\bibfnamefont{P.}~\bibnamefont{Weinberger}},
  \bibinfo{journal}{Phys. Rev. B} \textbf{\bibinfo{volume}{53}},
  \bibinfo{pages}{5125} (\bibinfo{year}{1996}).

\end{thebibliography}

\end{document}